# Large-scale Analysis of Opioid Poisoning Related Hospital Visits in New York State

Xin Chen, MS, Yu Wang, BS, Xiaxia Yu PhD, Elinor Schoenfeld, PhD, Mary Saltz, MD, Joel Saltz, MD, Fusheng Wang, PhD
Stony Brook University, Stony Brook, NY

**Abstract**

*Opioid related deaths are increasing dramatically in recent years, and opioid epidemic is worsening in the United States. Combating opioid epidemic becomes a high priority for both the U.S. government and local governments such as New York State. Analyzing patient level opioid related hospital visits provides a data driven approach to discover both spatial and temporal patterns and identity potential causes of opioid related deaths, which provides essential knowledge for governments on decision making. In this paper, we analyzed opioid poisoning related hospital visits using New York State SPARCS data, which provides diagnoses of patients in hospital visits. We identified all patients with primary diagnosis as opioid poisoning from 2010-2014 for our main studies, and from 2003-2014 for temporal trend studies. We performed demographical based studies, and summarized the historical trends of opioid poisoning. We used frequent item mining to find co-occurrences of diagnoses for possible causes of poisoning or effects from poisoning. We provided zip code level spatial analysis to detect local spatial clusters, and studied potential correlations between opioid poisoning and demographic and social-economic factors.*

**Introduction**

The United States is experiencing an epidemic of opioid related deaths. Since 2000, the rate of deaths from drug overdoses has increased 137%, including a 200% increase in the rate of overdose deaths involving opioids[1]. Given these alarming trends, combating opioid epidemic becomes a high priority for both the US government and local governments such as New York State. For example, the U.S. Department of Health and Human Services (HHS) has implemented evidence-based approaches to reduce opioid overdoses and the prevalence of opioid use disorder[2]. The MOON (Maximizing OpiOid safety with Naloxone) study tries to find out more information about the public's perception of opioid safety, naloxone distribution, and the use of the pharmacy as an integral site for public health intervention[3].

With increased accessibility of health data driven by open data initiatives, large-scale patient level data analysis provides an opportunity of data driven approach to identity patterns of opioid epidemic and discover potential causes of opioid related deaths through studying large scale diagnoses from hospital visits. This will provide quantitative measurements and essential knowledge to support the governments for decision making on prevention, treatment, recovery and enforcement. Analyzing opioid data at fine spatial resolutions such as zip code can also reveal community level distribution patterns in terms of demography, regions and historical trends, which will provide crucial knowledge for residents, schools, businesses and health and human service professionals for seeking solutions.

As part of New York State's open data initiative, New York State Statewide Planning and Research Cooperative System (SPARCS)[6] collects patient level details on patient characteristics, diagnoses and treatments, services, and charges for each inpatient stay and outpatient visit (emergency department, ambulatory surgery, and outpatient services). It also includes locations of patients (street addresses). Researchers can benefit from SPARCS data by leveraging patients' diagnosis histories, co-occurrences of diagnoses (primary and secondary ones), and location. Such unprecedented amounts of patient records make it possible for us to explore opioid poisoning in New York State with significant improvement of accuracy and coverage compared to prior studies on opioid abuse at global level[4] or urban level[5].

In this paper, we provide large scale analysis of opioid poisoning based on SPARCS data. We extract all patient records with opioid poisoning related primary diagnosis codes from all visits (inpatients, outpatients, emergency and ambulatory surgeries) from year 2010-2014 for our main studies, and from year 2003-2014 for temporal trend studies. We aim at four types of studies: 1) demographical based studies to explore disparities between population groups; 2) historical trends of opioid poisoning; 3) co-occurrence based studies to discover possible causes of opioid poisoning or effects from opioid poisoning; and 4) spatial analysis (zip code level) to explore potential local spatial clusters, and potential correlations between opioid poisoning and demographic and social-economic factors.

**Methods**

*Data Sources*

*SPARCS Data.* In this work, we used hospital discharge data from New York State SPARCS[6]. Any New York State healthcare facility certified to provide inpatient services, ambulatory surgery services, emergency department services or outpatient services is required to submit data to SPARCS. The purpose of SPARCS was to create a statewide data set to contribute to the goal of providing high quality medical care by serving as an information source[7].

*Opioid Poisoning Hospital Visits.* While the SPARCS data contains a comprehensive list of diagnosis and treatment procedure codes for each discharge record, this paper focused on analyzing patient level patterns of opioid poisoning hospital visits.

In SPARCS data, each hospital discharge record contains only one primary diagnosis code. Each record also contains one or more optional secondary diagnosis codes that include all conditions that coexisted at the time of admission, or developed subsequently, which affected the treatment received and/or length of stay[7].

In this work, we generated a subset of the SPARCS data using selected the International Classification of Diseases 9 (ICD-9) codes (also called billing codes) that pertain to opioid poisoning diagnoses, for both inpatient stays and outpatient visits (including emergency department visits, ambulatory surgery, and outpatient visits). We extracted opioid poisoning hospital visits by filtering the discharge records with their primary diagnosis code as the selected opioid poisoning ICD-9 codes.

The selected ICD-9 codes were a collection of poisonings by opiates, opium, heroin, methadone, and other related narcotics, including 9650 (Poisoning; Opiates and Related Narcotics), 96500 (Poisoning; Opium/alkaloids, unspecified), 96501 (Poisoning; Heroin), 96502 (Poisoning; Methadone), 96509 (Poisoning; Other opiates and related narcotics), E8500 (Accidental Poisoning; Heroin), E8501 (Accidental Poisoning; Methadone), and E8502 (Accidental Poisoning; Other Opiates and Related Narcotics).

**Table 1.** Demographics of patients with opioid related hospital visits, New York State, 2010-2014.

| | | New York State Population* | Patients with Opioid Poisoning Hospital Visits |
|---|---|---|---|
| | | 19,594,330 | 24,161 |
| Age and Sex | Under 5 years | 6.0% | 1.3% |
| | 5 to 14 years | 12.0% | 0.6% |
| | 15 to 44 years | 41.0% | 56.4% |
| | 45 to 64 years | 26.7% | 32.6% |
| | 65 and over | 13.5% | 9.0% |
| | Male persons | 48.5% | 60.0% |
| | Female persons | 51.5% | 40.0% |
| Race and Ethnic | White alone | 65.0% | 72.2% |
| | African American alone | 15.6% | 11.3% |
| | Asian alone | 7.8% | 0.5% |
| | Hispanic or Latino | 18.2% | 8.8% |

* 2010-2014 American Community Survey (ACS) 5-year estimates

*Patients' Residential ZIP Codes and Hospital Admission Years.* We conducted analyses at ZIP code level with basic demographics for the year 2010-2014 given in Table 1. We approximated patients' home location by combining the 5-digit ZIP code number with the patients' home address and geographic data from TIGER/LINE data[8]. We then aggregated opioid poisoning hospital visits at ZIP code level (Figure 1) and generated opioid poisoning incidence rates for the following spatial and temporal analyses (Figure 4-5).

For temporal trend studies, we used the hospital admission year for the time range 2005-2014 for inpatient stays and 2003-2014 for emergency department visits (Figure 3).

*Analysis Methods*

*Opioid Poisoning Incidence Rates*

We counted opioid poisoning patients with at least one opioid poisoning hospital at ZIP code level as shown in Figure 1. The opioid poisoning incidence rates by ZIP code were calculated through dividing counts of patients with opioid poisoning hospital visits by ZIP level population counts from Census data. In this paper, we evaluated both statewide rate and rates at ZIP code level. The incidence rates provided useful information about how common a disease is when compared to other diseases, or how common a disease in a specific location is as compared to the global baseline.

*Frequent Co-Occurrence Patterns*

Disease co-occurrence, also known as comorbidity[9-10] in public health studies, may imply the potential association across different types of diseases. The comorbidity is the presence of one or more additional diseases or disorders co-occurring with a primary disease or disorder[18].

We explored frequent disease co-occurrence patterns to find the frequent disease diagnoses that co-exist with opioid poisoning diagnoses. We first ranked all the secondary diagnosis codes by their counts in all the opioid poisoning hospital visits (Table 2). We then use Apriori-like algorithm[11] to discover top co-occurrences of diseases for opioid poisoning diagnoses (Table 3).

Apriori Algorithm is a common data mining technique to identify co-occurrences or temporal patterns between diseases in clinical domain[12]. This work tried to adopt Apriori algorithm to identify comorbidities among hospital visits. Apriori algorithm discovers frequent comorbidities by comparing their supports with a user-specified minimum support threshold.

For example, if the support of comorbidity pattern {96500 Poisoning by opium, 9670 Poisoning by barbiturates} (in Table 3) is 0.002, it means that 0.2% of all hospital visits have this comorbidity pattern. If the minimum support threshold is greater than 0.2%, this pattern will not be identified. However, if the minimum support threshold is set smaller than 0.2%, the comorbidity pattern will be extracted. In this work, we set the minimum support threshold as 0.1%.

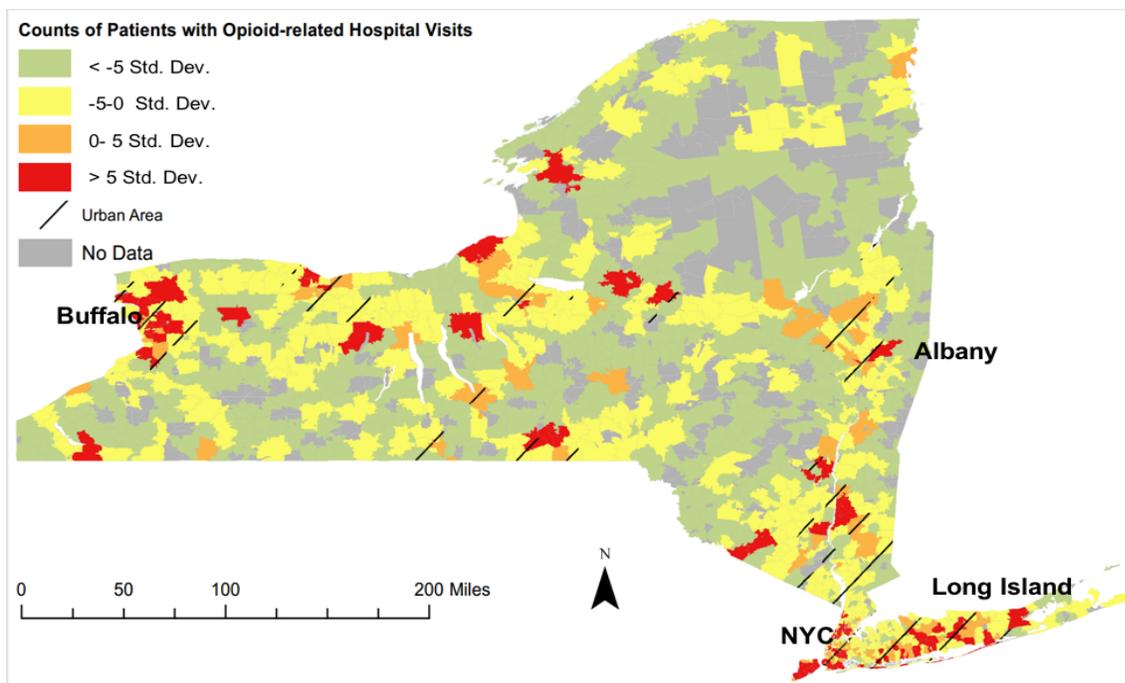

**Figure 1.** Counts of Patients with Opioid Poisoning Hospital Visits by Zip Code, New York State, 2010-2014.

*Spatial Clustering*

To test whether there is global spatial clustering tendency for opioid poisoning incidence rate or other spatial impact factors, we used Moran's I (Tables 4)[13]. Moran's I is a widely used global cluster test, which determines the degree of clustering or dispersion within a data set. The test result may range from 1 (perfect correlation), 0 (complete spatial randomness) to -1 (perfect dispersed). For the opioid poisoning incidence rate, a positive spatial autocorrelation means that the ZIP code areas with high rate are close to other areas with high rates.

In addition to the global cluster test (with Moran's I index in Table 4) and visual analysis for mapping raw counts of opioid poisoning incidence (in Figure 1), we also took cluster and outlier analysis with Anselin Local Moran's I statistics[13] to quantitatively detect local clusters for opioid poisoning incidence rates (Figure 4). The Anselin Local Moran's I statistics is a local cluster test that, given a set of weighted features, identifies statistically significant hot spots, cold spots, and spatial outliers.

*Spatial Regression*

To identify potential correlations between diseases and spatial impact indicators, we assessed both non-spatial and spatial correlation[13]. Ordinary least square regression analyses (OLS) was used to determined non-spatial correlation between opioid poisoning and demographic/socio-economic factors. We then used Geographically Weighted Regression (GWR) to assess the spatially varying relationship at a local level.

OLS is a linear regression method that closely fits a function by minimizing the sum of squared errors. To determine potential candidate factors, we evaluated several possible factor combinations that form a properly specified OLS regression model. GWR is a local form of linear regression used to model spatially varying relationships. To evaluate the correlation between opioid poisoning and spatial impact factors accounting for data in surrounding areas, we then conducted GWR with a selected spatial impact factor based on OLS results. Specifically, a fixed kernel type function was used to calculate the GWR regression coefficients. The extent of the kernel is determined using the Akaike Information Criterion (AICc)[14-15].

**Results**

*Demographic-based Analysis*

Table 1 compared the demographics between New York State population and patients with opioid poisoning during 2010-2014. The overall opioid poisoning incidence rate in New York State was 12.33 per 10,000 persons per five years. The percentage of female in New York State were 51.5% compared with only 40.0% of female patients with opioid poisoning, suggesting that female was less likely to have opioid poisoning hospital visits.

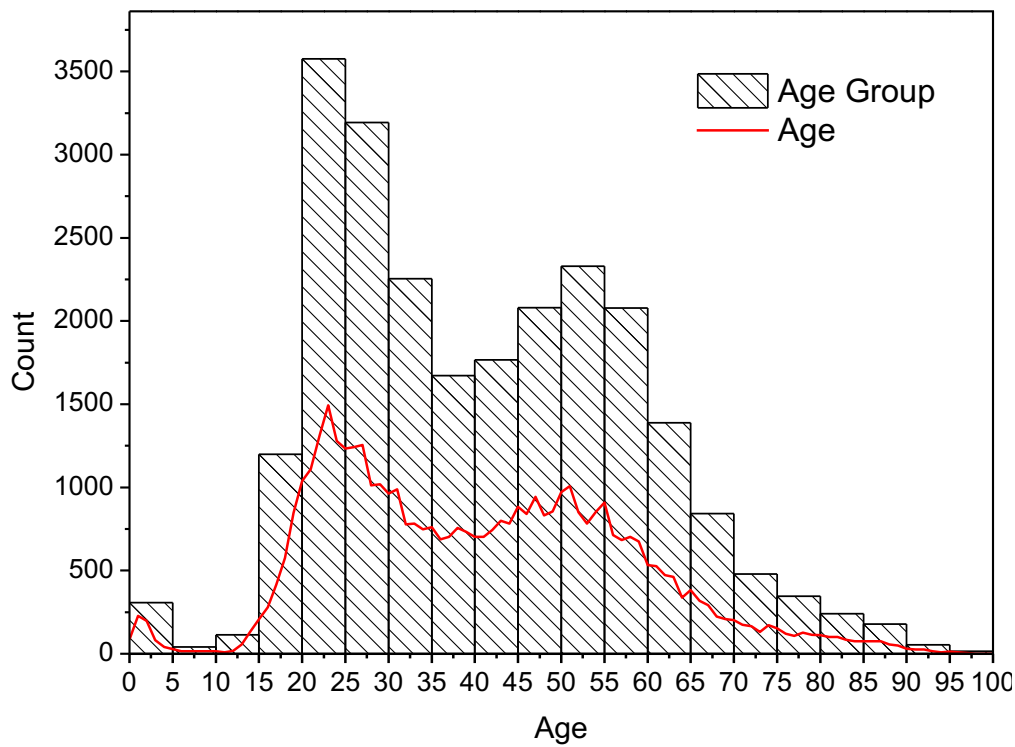

**Figure 2.** The age distribution of opioid poisoning in New York State, 2010-2014.

Comparing different race and ethnicity groups, we can see significant disparities between racial-ethnic proportions in patient and general population. Whites made 72.2% of patients that was higher than the whites proportion in general population. On the contrary, Asians had a very low percentage of patients that may require further investigation for the specific causes.

Figure 2 showed the age distribution of opioid poisoning in New York State, 2010-2014. We found a peak of opioid poisoning among young adults aged 21 to 25. Prior work has shown an increasing trend of adolescents and young adults with prescription drugs misuse[16-17]. Our statistics on age for opioid poisoning hospital visits (Figure 2) was consistent with the prior national-wide survey.

*Temporal Trends*

Outpatient emergency departments (EDs) and inpatient stays play an important role in the treatment of drug poisoning. For the period of most recent year available 2010-2014, both opioid poisoning hospital EDs and inpatient stays continued to increase in the United States, as is true for New York State from our results. Statewide, there were 4,238 outpatient ED visits and 2,909 inpatient stays in 2014, a 116.4 percent increase and a 17.3 percent increase respectively from 2010.

Figure 3 showed temporal trends of the proportion of opioid poisoning hospital visits for the past decade. While we found a modest rising trend for inpatient stays, there was an increasingly rising trend for outpatient EDs.

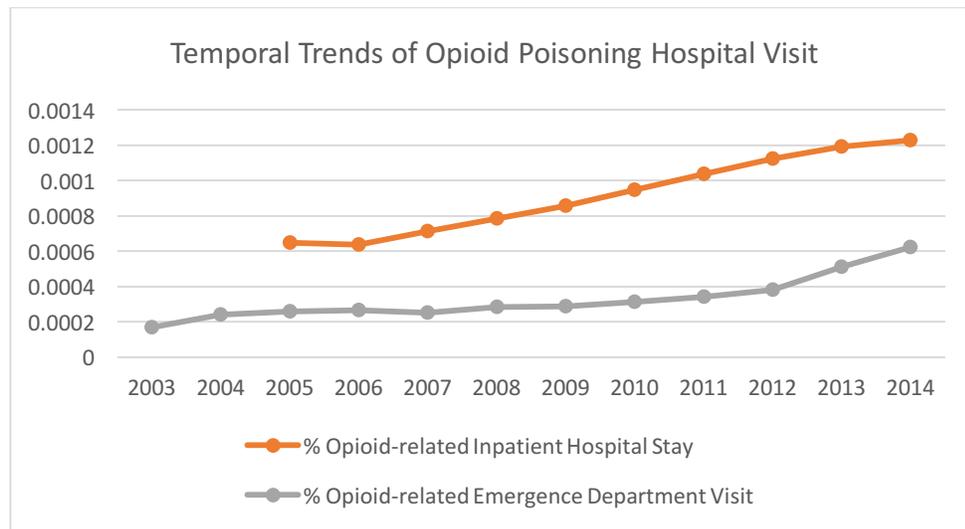

**Figure 3.** Temporal trends of the proportion of opioid poisoning hospital visits, New York State, 2003-2014.

*Frequent Co-Occurrence Patterns*

We used frequent item mining to find co-occurrences of diagnoses for possible causes of poisoning or effects from poisoning. We used all ICD-9 diagnosis codes from opioid poisoning hospital discharge records. The diagnosis codes mark medical conditions, such as chronic ischemic heart disease, pure hypercholesterolemia, type 2 diabetes, and many other types of diseases.

Table 2 presented the top 20 diagnosis codes that co-existed with opioid poisoning diagnoses by their total counts in all opioid poisoning discharge records.

We found several diagnoses that were the well-established comorbidities of opioid dependence. For example, opioid use disorder is often associated with other substance use disorders, such as tobacco, alcohol and benzodiazepines, which are often taken to reduce symptoms of opioid withdrawal or craving for opioids, or to enhance the effects of administered opioids. Individuals with opioid use disorder are also at risk for the development of mild to moderate depression. Periods of depression are especially common during chronic intoxication or in association with physical or psychosocial stressors that are related to the opioid use disorder.[18]

As we didn't find any infection disease among the top co-existed diagnoses in Table 2, most of the opioid poisoning hospital visits may come from patients with prescription opioids. Usually, the most common medical conditions associated with opioid use disorder are viral (e.g., HIV, hepatitis C virus) and bacterial infections, particularly among users of opioids by injection. These infections are less common in opioid use disorder with prescription opioids.[18]

The other co-existed diagnoses listed in Table 2 should be the effects from opioid poisoning, including acute respiratory failure, acute kidney failure, pneumonitis, hypopotassemia, and rhabdomyolysis.

**Table 2.** Top 20 Co-existed Diagnosis Codes of Opioid Poisoning Hospital Visits.

| Code | Description | Rank | Code | Description | Rank |
|---|---|---|---|---|---|
| 3051 | Tobacco Use Disorder | 1 | 5849 | Acute Kidney Failure, Unspecified | 11 |
| 4019 | Unspecified Essential Hypertension | 2 | 25000 | Diabetes Mellitus Without Mention Of Complication | 12 |
| 311 | Depressive Disorder, Not Elsewhere Classified | 3 | 30000 | Anxiety State, Unspecified | 13 |
| 30550 | Opioid Abuse, Unspecified | 4 | 5070 | Pneumonitis Due To Inhalation Of Food Or Vomitus | 14 |
| 51881 | Acute Respiratory Failure | 5 | 30500 | Alcohol Abuse, Unspecified | 15 |
| 30401 | Opioid Type Dependence, Continuous | 6 | 30590 | Other, Mixed, Or Unspecified Drug Abuse, Unspecified | 16 |
| 9694 | Poisoning By Benzodiazepine-Based Tranquilizers | 7 | 2768 | Hypopotassemia | 17 |
| 78097 | Altered Mental Status | 8 | 33829 | Other Chronic Pain | 18 |
| 78009 | Other Alteration Of Consciousness | 9 | 49390 | Asthma, Unspecified Type, Unspecified | 19 |
| 30400 | Opioid Type Dependence, Unspecified | 10 | 72888 | Rhabdomyolysis | 20 |

**Table 3.** Top Co-existed diagnosis codes for different types of opiates and related narcotics.

| Opioid poisoning Diagnosis Code and Description | Co-existed Diagnosis | | |
|---|---|---|---|
| | Code | Description | Support |
| 96500 Poisoning; Opium (alkaloids) | 9670 | Poisoning By Barbiturates | 0.002 |
| | 9679 | Poisoning By Unspecified Sedative Or Hypnotic | 0.001 |
| 96501 Poisoning; Heroin | 30550 | Opioid Abuse, Unspecified | 0.074 |
| | 30551 | Opioid Abuse, Continuous | 0.012 |
| | 97081 | Poisoning By Cocaine | 0.009 |
| | 9708 | Poisoning by other specified central nervous system stimulants | 0.009 |
| | 4275 | Cardiac Arrest | 0.006 |
| 96502 Poisoning; Methadone | E8501 | Accidental Poisoning; Methadone | 0.002 |
| 96509 Poisoning; Other opiates and related narcotics | 33829 | Other Chronic Pain | 0.027 |
| | 7245 | Backache, Unspecified | 0.019 |
| | 9654 | Poisoning By Aromatic Analgesics | 0.013 |
| | 7242 | Lumbago | 0.013 |
| | 2720 | Pure Hypercholesterolemia | 0.009 |

Table 3 presented the top co-existed diagnoses for poisoning by different substance. Poisoning by heroin should be mainly caused by drug abuse as it generally co-existed with opioid abuse and resulted in a life-threating cardiac arrest. Poisoning by opium more likely co-existed with poisoning by barbiturates or other sedative-hypnotic drugs. On the other hand, poisoning by methadone mainly co-existed with accidental poisoning by methadone. Such findings may require further research for the potential driving factors.

*Spatial Analysis*

In this section, we performed zip code level spatial analysis to detect local spatial clusters, and studied potential correlations between opioid poisoning and demographic and social-economic factors.

Figure 4 demonstrated the results of spatial clusters and outliers. The cluster/outlier type field distinguishes between a statistically significant cluster of high values (High-High cluster), cluster of low values (Low-Low cluster), outlier in which a high value is surrounded by low values (High-Low outlier), and outlier in which a low value is surrounded by high values (Low-High outlier). Statistical significance is set at the 95 percent confidence level. We applied the False Discovery Rate (FDR) correction to reduce this p-value threshold from 0.05 to a value that better reflects the 95 percent confidence level given multiple testing. The FDR procedure will potentially reduce the critical p-value to account for multiple testing and spatial dependency.[13]

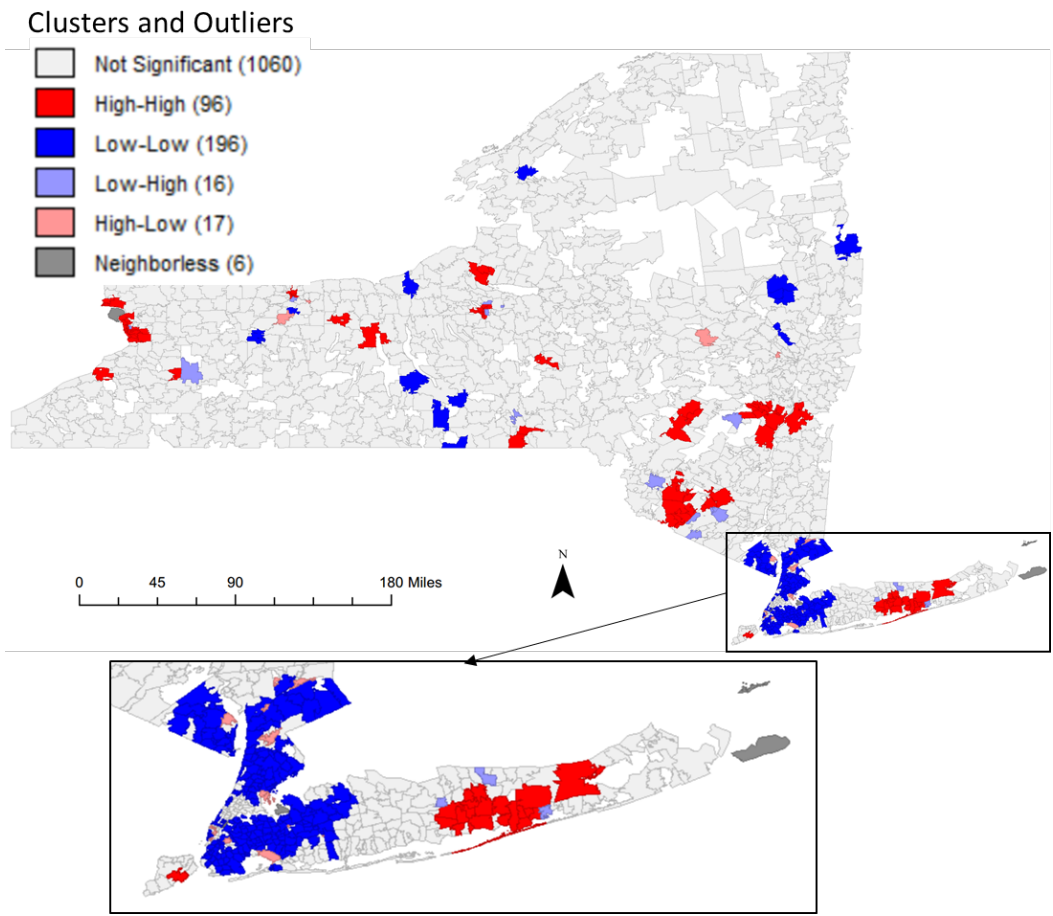

**Figure 4.** Spatial clusters and outliers of opioid poisoning incidence rate by ZIP code, New York State, 2010-2014.

We found that most areas of New York City had lower opioid poisoning hospital visits (Low-Low clusters in blue color) except several scattered high risk areas. Such findings were consistent with the decreasing trend of opioid overdose[19] in New York City and the steep rise in opioid hospital visits outside New York City[20]. Targeted public health interventions might be effective in lowering opioid overdose mortality rates[19]. However, a small part of Staten Island and most part of southeastern Long Island were identified as high risk areas (High-High clusters in red color).

For the scattered outliers across New York City (High-Low outlier or Low-High outlier), further investigation and intervention efforts may be required in the future.

The spatial clustering analysis tried to answer the question of "where" by identifying potential high or low risk areas. The following question should be finding out why there were higher or lower opioid poisoning hospital visits at the discovered spatial clusters or outliers. We then used spatial regression analysis by linking demographic and socio-economic factors to opioid poisoning.

We did exploratory regression analyses between the opioid poising and exemplary demographic and socio-economic factors as shown in Table 4. Among the 13 spatial impact factors, the OLS regression results showed that household income, Asian population and the youth population age 20 to 24 had most significant negative correlation relationship with the opioid poisoning.

We then used GWR to model the local trends of correlation relationship between household income and opioid poisoning rates. The local coefficient map in Figure 5 showed that urban residents who made more money had lower rates of opioid poisoning, while country residents with higher income tended to had higher rates of opioid poisoning.

**Table 4.** Statistics of demographic and socio-economic factors by ZIP Code in New York State, 2010-2014.

| Spatial Impact Factors | | Mean (Std. Error) | Moran's I Index |
|---|---|---|---|
| Opioid Poisoning Incidence Rate per 1,000 | | 1.68 (2.02) | 0.09 |
| Demographic Factors | % Male | 49.61 (5.41) | 0.07 |
| | % Age 15-19 years | 6.73 (4.32) | 0.07 |
| | % Age 20-24 years | 6.47 (4.06) | 0.14 |
| | % Age 25-34 years | 11.71 (5.39) | 0.44 |
| | % Age 35-44 years | 12.27 (3.74) | 0.13 |
| | % White | 78.25 (25.22) | 0.73 |
| | % Black | 6.89 (14.14) | 0.63 |
| | % Asian | 3.69 (7.01) | 0.72 |
| | % Hispanic | 8.90 (12.63) | 0.69 |
| Socio-economic Factors | Household Income ($10,000 US) | 76,934 (35,610) | 0.63 |
| | % Poverty | 12.5 (10.01) | 0.43 |
| | % Uninsured | 9.15 (5.59) | 0.30 |

**Discussion**

There has long been a demand for large-scale data driven approach to discover both spatial and temporal patterns and identity potential causes of diseases. This study provided our preliminary results of a large-scale patient level study on opioid poisoning hospital visits in the New York State. We examined the demographic disparities for the patients with opioid poisoning. We compared the historical trends of opioid poisoning for hospital emergency departments visits and inpatient stays. We used frequent item mining to find co-occurrences of diagnoses for possible causes of poisoning or effects from poisoning. We performed zip code level spatial analysis to detect local spatial clusters, and studied potential correlations between opioid poisoning and demographic and social-economic factors.

This work focuses on a large-scale study of opioid poisoning related hospital visits in New York State. We found that men were more likely to have opioid poisoning hospital visits than women. Whites made 72.2% of patients that was higher than the whites proportion in general population. Asians, however, had a very low percentage of patients. We also found a peak of opioid poisoning hospital visits for young adults (21 to 25, peak at 22). A second peak is for ages starting from middle age (46-55, peak at 51). We found several diagnoses that were well-established comorbidities of opioid dependence. The spatial clustering analysis showed that most areas of New York City had lower opioid poisoning hospital visit rates compared to that in areas outside New York City. The spatial regression results showed that urban residents who made more money had lower rates of opioid poisoning, while country residents with higher income tended to had higher rates of opioid poisoning.

However, our results were based on the patients with their primary diagnosis as opioid poisoning. In our ongoing work, we will also include the patients with their secondary diagnosis as opioid poisoning. We will examine more types of disease diagnoses and treatments and provide multi-dimensional analysis by grouping patients per their demographic attributes. We will take advantage of the street level location information and the full history of each patient for more fine grained spatial and temporal analysis. For example, after geocoding the patient addresses into latitude and longitude coordinates, we will perform point based spatial clustering analysis. By mapping patients' residential locations into census block group boundaries, we will link the health records with census data and study potential risk factors with fine-grained spatial resolutions.

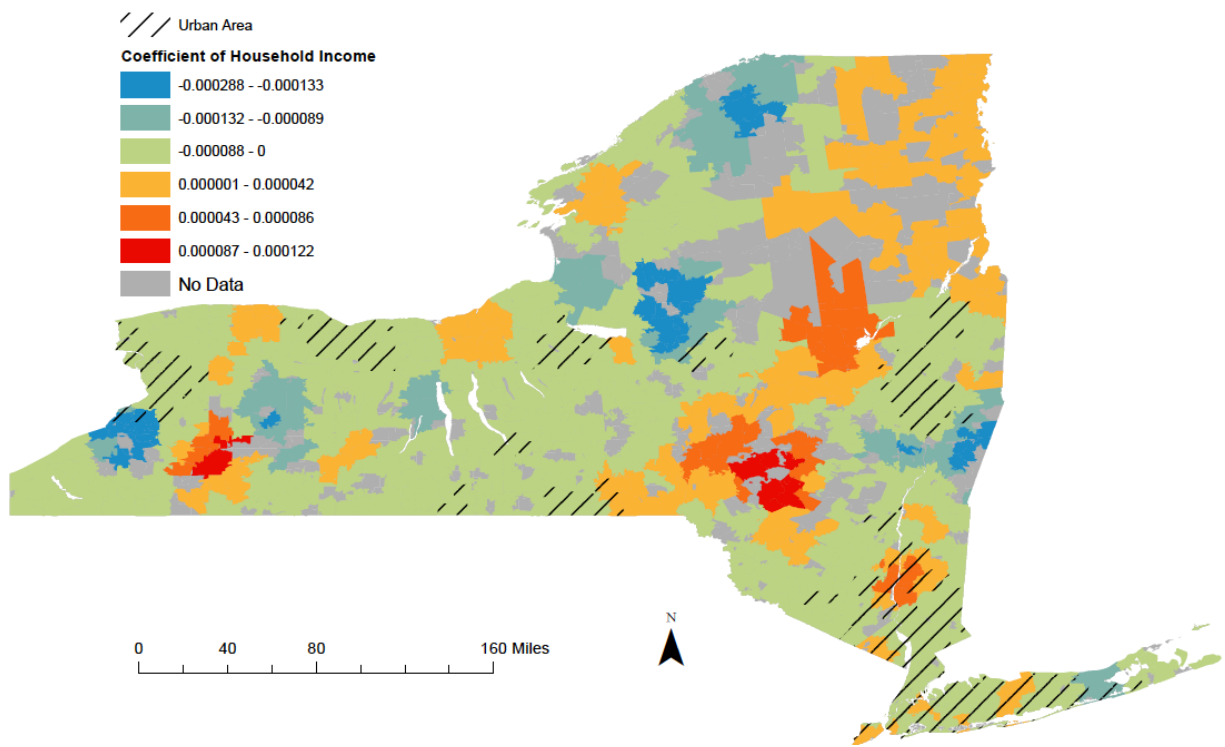

**Figure 5.** The choropleth map that visualizes Geographically Weighted Regression (GWR) local coefficient of household Income for opioid poisoning incidence rate by ZIP code in New York State, 2010-2014.

**Conclusion**

Increased accessibility of health data made available by the government provides unique opportunity for data driven discovery of disease patterns and identity potential causes of opioid related deaths. large-scale patient level analysis could provide new insights and create new forms of value to support governments on decision making. In this paper, we present our analytical results on opioid poisoning related hospital visits using New York State SPARCS data. Our results not only provide quantitative measurements on demographic-based distributions, and spatial and temporal trends, but also present top common co-occurrences of diagnoses with opioid poisoning, which provides a groundwork to study possible causes of poisoning and effects from poisoning. Our results provide essential knowledge and guidelines to support stake holders for improving prevention, intervention and recovery of opioid poisoning.

**Acknowledgments**

This work is supported in part by NSF ACI 1443054 and by NSF IIS 1350885.